\documentclass[12pt]{article}
\usepackage{charter} % change font
\usepackage{fullpage}
\usepackage{titlesec}
\usepackage[hyphens]{url}     % allow hyphens in the url
\usepackage[hidelinks]{hyperref}       % hyperlinks
\usepackage{booktabs}       % professional-quality tables
\usepackage{amsfonts}       % blackboard math symbols
\usepackage{amssymb}        % varnothing
\usepackage{nicefrac}       % compact symbols for 1/2, etc.
\usepackage{microtype}      % microtypography
\usepackage{bm}				% italic bold
\usepackage{cite}			% group citations
\usepackage{graphicx}		% figures
\usepackage{mathtools}		% for "coloneqq"
\usepackage{algorithm}
\usepackage{algpseudocode}
\usepackage{amsthm}			% theorems
\usepackage{enumitem}		% to have letter enumerations
\usepackage{multirow}       % to generate tables
\usepackage{tabularx}	    % to generate tables
\usepackage{hhline}
\usepackage{arydshln}		% dotted lines in a table
\usepackage{xcolor}
\usepackage{siunitx}

\definecolor{lightgreen}{rgb}{.9,1,.9}

\newcolumntype{L}[1]{>{\raggedright\arraybackslash}p{#1}}
\newcolumntype{C}[1]{>{\centering\arraybackslash}p{#1}}
\newcolumntype{R}[1]{>{\raggedleft\arraybackslash}p{#1}}

\theoremstyle{plain} % plain = italic, definition = roman

\def\defn{\,\coloneqq\,}
\def\argmin{\mathop{\mathsf{arg\,min}}} % Argument of a minimization

\def\lim{\mathop{\mathsf{lim}}} % limit

\def\prox{\mathsf{prox}}

\def\Dcal{\mathsf{\, d}}

\def\proposed{CoIL}

\def\ebm{{\bm{e}}}

\def\sbm{{\bm{s}}}

\def\xbm{{\bm{x}}}

\def\xbmtilde{{\tilde{\xbm}}}

\def\ybm{{\bm{y}}}
\def\zbm{{\bm{z}}}

\def\vbm{{\bm{v}}}

\def\Abm{{\bm{A}}}
\def\Dbm{{\bm{D}}}

\def\xbmhat{{\widehat{\bm{x}}}}
\def\zbmhat{{\widehat{\bm{z}}}}

\def\Tsf{{\mathsf{T}}}

\def\R{\mathbb{R}}

\def\Mcal{{\mathcal{M}}}
\def\Ncal{{\mathcal{N}}}
\def\Hcal{{\mathcal{H}}}
\def\Gcal{{\mathcal{G}}}
\def\Dcal{{\mathcal{D}}}

\def\Lcal{{\mathcal{L}}}
\def\Fcal{{\mathcal{F}}}

\def\sin{{\textit{sin}}}
\def\cos{{\textit{cos}}}

\begin{document}

\title{CoIL: Coordinate-based Internal Learning for Imaging Inverse Problems}

%\author{Yu~Sun,~\IEEEmembership{Student~Member,~IEEE}, Jiaming Liu,~\IEEEmembership{Student~Member,~IEEE}, Mingyang Xie, \\ Brendt Wohlberg,~\IEEEmembership{Senior Member,~IEEE}, and Ulugbek~S.~Kamilov,~\IEEEmembership{Senior Member,~IEEE}%
%\thanks{This work was supported by NSF award CCF-1813910 and by the Laboratory Directed Research and Development program of Los Alamos National Laboratory under project number 20200061DR. \emph{(Corresponding author: Ulugbek~S.~Kamilov.)}}
%\thanks{Y.~Sun and M.~Xie is with the Department of
%Computer Science \& Enginnering, Washington University in St.~Louis, MO 63130, USA.}
%\thanks{Jiaming Liu is with the Department of Electrical \& Systems Engineering, Washington University in St.~Louis, MO 63130, USA.}
%\thanks{B.~Wohlberg is with Theoretical Division, Los Alamos National Laboratory, Los Alamos, NM 87545 USA.}
%\thanks{U.~S.~Kamilov (email:~kamilov@wustl.edu) is with the Department of Computer Science \& Engineering and the Department of Electrical \& Systems Engineering, Washington University in St.~Louis, MO 63130, USA.}}
%
%\markboth{CoIL: Coordinate-based Internal Learning for Imaging Inverse Problems}%
%{Sun,~Liu,~Xie~and~Kamilov}

\author{
Yu~Sun$^{\footnotesize 1}$, Jiaming Liu$^{\footnotesize 2}$, Mingyang Xie$^{\footnotesize 1}$, \\ 
Brendt~Wohlberg$^{\footnotesize 3}$,~and~Ulugbek~S.~Kamilov$^{\footnotesize 1, 2, \ast}$\\
\emph{\footnotesize $^{\footnotesize 1}$Department of Computer Science and Engineering,~Washington University in St.~Louis, MO 63130, USA}\\
\emph{\footnotesize $^{\footnotesize 2}$Department of Electrical and Systems Engineering,~Washington University in St.~Louis, MO 63130, USA}\\
\emph{\footnotesize $^{\footnotesize 3}$Los Alamos National Laboratory, Theoretical Division, Los Alamos, NM 87545 USA}\\
\small$^{\footnotesize *}$\emph{Email}: \texttt{kamilov@wustl.edu}
}

\date{}

\maketitle %% required

\begin{abstract}
We propose \emph{Coordinate-based Internal Learning (\proposed)} as a new deep-learning (DL) methodology for continuous representation of measurements.
Unlike traditional DL methods that learn a mapping from the measurements to the desired image, \proposed~trains a multilayer perceptron (MLP) to encode the complete measurement field by mapping the coordinates of the measurements to their responses. \proposed~is a self-supervised method that requires no training examples besides the measurements of the test object itself.  Once the MLP is trained, \proposed~generates new measurements that can be used within a majority of image reconstruction methods. We validate \proposed~on sparse-view computed tomography using several widely-used reconstruction methods, including purely model-based methods and those based on DL. Our results demonstrate the ability of \proposed~to consistently improve the performance of all the considered methods by providing high-fidelity measurement fields.
\end{abstract}

\section{Introduction}

The problem of reconstructing an unknown image from a set of noisy measurements is fundamental to computational imaging. The task is traditionally formulated as an inverse problem and solved using model-based optimization by leveraging a forward model characterizing the imaging system and a regularizer imposing prior knowledge on the unknown image. There has been significant progress in developing sophisticated image priors, including those based on transform-domain sparsity, self-similarity, and dictionary learning~\cite{Rudin.etal1992, Figueiredo.Nowak2001, Elad.Aharon2006, Danielyan.etal2012}.

There has been a considerable recent interest in \emph{deep learning (DL)} based solutions to imaging inverse problems~\cite{McCann.etal2017, Lucas.etal2018, Ongie.etal2020, Wang.etal2020}. The traditional DL approach involves training a convolutional neural network (CNN) to directly perform a regularized inversion of the forward model by exploiting redundancies in a training dataset~\cite{DJin.etal2017, Kang.etal2017, Sun.etal2018, han.etal2018}. Model-based DL is an alternative to the traditional DL that explicitly uses knowledge of the forward model by integrating a CNN into model-based optimization. Two widely-used approaches in this context are \emph{plug-and-play priors (PnP)}~\cite{Venkatakrishnan.etal2013} and \emph{regularization by denoising (RED)}~\cite{Romano.etal2017}, which have been used with pre-trained deep denoisers to achieve excellent performance in a number of imaging tasks~\cite{Ono2017, Kamilov.etal2017, Bigdeli.etal2017, Wu.etal2020, Liu.etal2020, Chan.etal2016, Buzzard.etal2017, Ryu.etal2019, Sun.etal2019a,Xu.etal2020, Sun.etal2019c}. An alternative model-based DL approach is \emph{deep unfolding}, which interprets the iterations of a model-based optimization algorithm as layers of a CNN and trains it end-to-end in a supervised fashion~\cite{zhang2018ista, Yang.etal2016, Hauptmann.etal2018, Adler.etal2018, Aggarwal.etal2019, Hosseini.etal2019, Chun.etal2020, Yaman.etal2020, Aggarwal.etal2020, Kellman.etal2020, Liu.etal2021}.

%%%%%%%%%%%%%%%%%%%%%%%%%%%
\begin{figure}[t!]
\begin{center}
\includegraphics[width=0.4\linewidth]{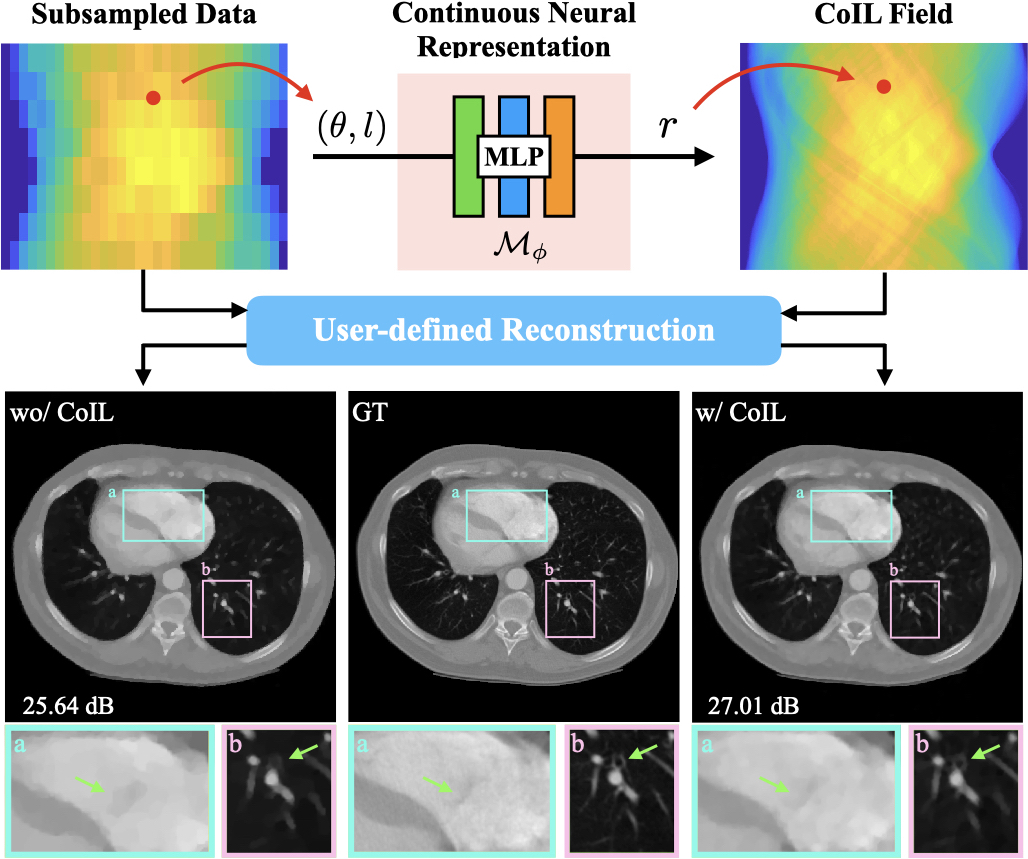}
\end{center}
\caption{The conceptual illustration of \proposed~in the context of sparse-view CT. A multilayer perceptron (MLP) is used to represent the full measurement field by learning to map the measurement coordinate $(\theta, l)$ to its response $r$. Visual examples compare the recovered images with and without \proposed~for total variation (TV). \proposed~is used to generate $360$ views from the data consisting of $120$ noisy views of $40$ dB input SNR. The quantitative and visual results in this paper highlight the ability of \proposed~to significantly improve the imaging quality for several widely-used image reconstruction methods.}
\label{Fig:banner}
\end{figure}
%%%%%%%%%%%%%%%%%%%%%%%%%%%

There has been a considerable amount of work on DL for imaging inverse problems, the unifying theme being that one can train a CNN over a dataset to represent a prior for an unknown image. In this paper, we take a fundamentally different approach by proposing a methodology for leveraging redundancy within the measurements of a single unknown image, thus requiring no training examples besides the test-input itself. Our proposed \emph{Coordinate-based Internal Learning (\proposed)} seeks to represent the full continuous measurement field by exploiting the internal information within the subsampled and noisy measurements. The core of \proposed~is a \emph{multilayer perceptron (MLP)} that maps the measurement coordinates to the corresponding sensor responses.
The measurement coordinates are the parameters corresponding to the geometry of the imaging system that determine the response measured by the sensors. For example, in computed tomography (CT) two parameters characterizing the response are the view angle $\theta$ of the incoming ray beam and the spatial location $l$ of the relevant detector on the sensor plane.
By training MLP on the coordinate-response pairs extracted from the measurements of a desired object, \proposed~is able to build a continuous mapping from the coordinates to the sensor responses.
Thus, the learned MLP corresponds to a neural representation of the full measurement field. By querying the MLP with the relevant coordinates, \proposed~can generate the full field that can be used for image reconstruction.  Figure~\ref{Fig:banner} provides a conceptual illustration of the \proposed~methodology.
Note that \proposed~is not restricted to a specific image reconstruction method, but is compatible with a majority of methods including those based on model-based optimization or DL.

The main contributions of this paper are as follows:
\begin{itemize}
\item We propose \proposed~as the first computational imaging methodology that leverages a coordinate-based neural representation~\cite{Mildenhall.etal2020, Martin.etal2020, Zhang.etal2020} for learning high-quality measurement fields. Our work complements the recent work in DL by exploiting the internal information in the measurements, which can be subsequently combined with other information sources during reconstruction.

\item We propose a novel MLP architecture for representing the measurement field. Unlike the CNN architectures that rely on a sequence of multi-filter convolutions, the MLP is built on  fully-connected layers with only $256$ hidden neurons. The relatively small scale of our model makes it straightforward to train and deploy.

\item We extensively validate our method in the context of sparse-view CT. We show that \proposed~synergistically combines with a majority of widely-used methods by being able to generate high-fidelity full-view sinograms. In all our experiments the methods with \proposed~consistently outperform the corresponding ones without it.
\end{itemize}

%The rest of the paper is organized as follows.
%Section~\ref{Sec:Background} reviews the background on computational imaging and coordinate-based learning.
%Section~\ref{Sec:Proposed} describes the technical details of the \proposed~framework.
%Section~\ref{Sec:Experiments} presents the numerical validation on sparse-view CT.
%Section~\ref{Sec:Conclusion} concludes the paper.

%%%%%%%%%%%%%%%%%%%%%%%%%%%
\begin{figure*}[t]
\begin{center}
\includegraphics[width=0.95\linewidth]{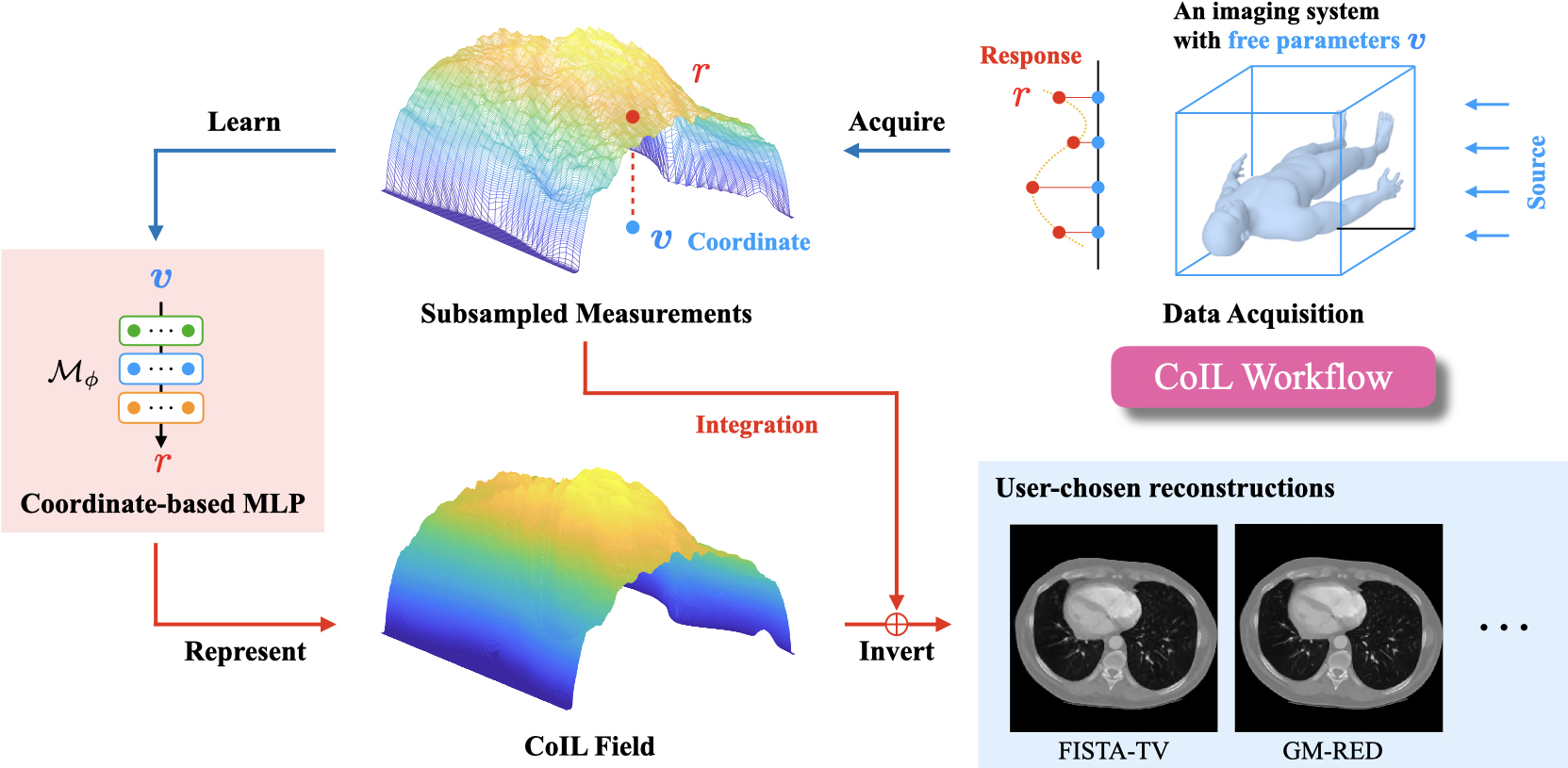}
\end{center}
\caption{Illustration of the \proposed~workflow for an arbitrary imaging system with  free parameters $\vbm\in\R^v$.
First, a set of $N>0$ measurements are acquired by the system under different realization of $\vbm$.
Then, the coordinate-response pairs $\{(\vbm_i,r_i)\}_{i=1}^{N}$ are used to train a coordinate-based MLP $\Mcal_\phi:\vbm\rightarrow r$ for encoding the full measurement field. Once the training is finished, the encoded field is extracted from $\Mcal_\phi$ with an arbitrary resolution by querying the relevant coordinates. In the final stage, the \proposed~field and the actual measurements are jointly used for image reconstruction using a user-defined method.}
\label{Fig:workflow}
\end{figure*}
%%%%%%%%%%%%%%%%%%%%%%%%%%%

\section{Background}
\label{Sec:Background}
In this section, we review background information related to \proposed.
We introduce the imaging inverse problems and review several popular reconstruction methods.
We also discuss sensor-domain DL models and the recent progress on internal learning.

\subsection{The inverse problem in imaging}
Consider the linear measurement model
\begin{equation}
\label{Eq:inverseProblem}
\ybm = \Abm\xbm+ \ebm\,,
\end{equation}
where the measurement operator $\Abm \in \R^{m \times n}$ characterizes the response of the imaging system and vector $\ebm \in \R^m$ represents the noise, which is often assumed to be additive white Gaussian (AWGN).
The associated imaging inverse problem involves the reconstruction of the image $\xbm\in\R^n$ from the measurements $\ybm\in\R^m$.
%When the inverse problem is nonlinear, the forward operator can be generalized to a more general mapping ${\Hbm: \R^{n} \rightarrow \R^m}$ with $\ybm = \Hbm(\xbm) + \ebm$.
Due to ill-posedness, practical inverse problems are often formulated as regularized optimization
\begin{equation}
\label{Eq:Optimization}
\widehat{\xbm} = \argmin_{\xbm \in \R^{n}}\; f(\xbm), \quad\text{with}\quad f(\xbm)=g(\xbm) + h(\xbm),
\end{equation}
where $g$ is the data-fidelity term that quantifies the consistency of $\xbm$ with $\ybm$, and $h$ is the regularizer that imposes some prior knowledge on $\xbm$.
%\textcolor{red}{Describe FBP}
For instance, two widely used functions in the context of imaging are least-square and total variation (TV)
\begin{equation}
\label{Eq:LeastSquares}
g(\xbm) = \frac{1}{2}\|\Abm\xbm-\ybm\|_2^2 \quad\text{and}\quad h(\xbm) = \tau\|\Dbm\xbm\|_{1},
\end{equation}
where $\tau > 0$ controls the regularization strength and $\Dbm$ is the discrete gradient operator~\cite{Rudin.etal1992}. The nonsmoothness of the regularizer is a common occurrence in imaging, which precludes the usage of the standard gradient descent algorithms.

The family of proximal methods are effective solvers for nonsmooth optimization problems of form~\eqref{Eq:Optimization}.
Two common algorithms are \emph{fast iterative shrinkage/thresholding algorithm} (FISTA)~\cite{Beck.Teboulle2009} and \emph{alternating direction method of multipliers} (ADMM)~\cite{Boyd.etal2011}.
These algorithms rely on a mathematical concept known as the \emph{proximal operator}~\cite{Moreau1965}, defined as
\begin{equation}
\label{Eq:ProximalOperator}
\prox_{\mu h}(\zbm) \defn \argmin_{\xbm \in \R^n} \left\{\frac{1}{2}\|\xbm-\zbm\|_2^2 + \mu h(\xbm)\right\},
\end{equation}
to handle nonsmooth terms without differentiation. The parameter $\mu>0$ in~\eqref{Eq:ProximalOperator} balances the importance of the term $h$. Note that the proximal operator can be interpreted as a \emph{maximum a posterior} (MAP) denoiser for AWGN with variance $\mu$.

\subsection{Traditional deep learning methods}
\label{Sec:EndToEnd}

DL has become very popular for imaging inverse problems~\cite{McCann.etal2017, Lucas.etal2018, Ongie.etal2020, Wang.etal2020} due to its excellent performance. 
Traditional DL methods first bring the measurements $\{\ybm_i\}_{i=1}^N$ to the image domain and then use a deep CNN architecture, such as UNet~\cite{Ronneberger.etal2015}, to map the resulting low-quality images $\{\xbmtilde_i\}_{i=1}^N$ to high-quality images $\{\xbm_i\}_{i=1}^N$.
Here, $N>0$ denotes the total number of training examples.
Typically, these CNNs are trained by minimizing a loss function
\begin{equation}
\label{Eq:CNNLoss}
\ell(\psi)=\frac{1}{N}\sum_{i=1}^{N} \Lcal (\Fcal_{\psi}(\xbmtilde_{i}),\xbm_{i}),
\end{equation}
where $\Fcal_{\psi}$ denotes the network parametrized by $\psi$, and function $\Lcal$ quantifies the discrepancy between $\Fcal_{\psi}(\xbmtilde_{i})$ and $\xbm_{i}$.
Popular choices for $\Lcal$ include the $\ell_1$ and $\ell_2$ norms.
Some other models consider different schemes that directly map $\{\ybm_i\}$ to reconstructed images $\{\xbm_i\}$.
These methods often adopt hybrid CNN architectures that contain fully connected layers for learning either an approximation of the inverse $(\Abm\Abm^\Tsf)^{-1}$~\cite{Wurfl.etal2018} or an inversion to some implicit image manifolds~\cite{Zhu.etal2018}.
Nevertheless, traditional DL methods do not explicitly impose consistency with respect to the forward model during image reconstruction.

\subsection{Deep denoising priors}
The family of denoising-driven approaches represents an alternative to traditional DL by combining iterative model-based algorithms with deep denoisers as priors.
These methods draw inspiration from the equivalence between proximal operator and image denoiser.
One popular framework is PnP, which generalizes the proximal methods, such as FISTA and ADMM, by replacing the proximal operator with an arbitrary AWGN denoiser $\Dcal_\sigma:\R^n\rightarrow\R^n$, with $\sigma>0$ controling the denoising strength.
This simple replacement enables PnP to use advanced denoisers, including those based on CNNs~\cite{Zhang.etal2017,Zhang.etal2018,Song.etal2020}, for regularizing the inverse problem.
The PnP algorithms have been shown to be effective in various imaging applications~\cite{Sreehari.etal2016, Zhang.etal2017a, Sun.etal2019b, Zhang.etal2019, Ahmad.etal2020}.
However, $\Dcal_\sigma$ may not correspond to any explicit $h$ in~\eqref{Eq:Optimization}, in which case PnP loses its interpretation as optimization. PnP has also been theoretically analyzed PnP~\cite{Chan.etal2016, Meinhardt.etal2017, Buzzard.etal2017, Sun.etal2019a, Tirer.Giryes2019, Teodoro.etal2019, Ryu.etal2019, Xu.etal2020}.

RED~\cite{Romano.etal2017} is a related framework that uses the operator
\begin{equation}
\Hcal(\xbm) = \tau(\xbm - \Dcal_\sigma(\xbm)),
\end{equation}
within many kinds of iterative algorithms~\cite{Romano.etal2017,Sun.etal2019c,Wu.etal2020,Sun.etal2021}.
RED with deep denoisers has been reported to be effective in image super-resolution~\cite{Mataev.etal2019}, phase retrieval~\cite{Metzler.etal2018}, and tomographic imaging~\cite{Wu.etal2020}.
It has been shown that when the denoiser $\Dcal_\sigma$ is locally homogeneous and has a symmetric Jacobian~\cite{Romano.etal2017, Reehorst.Schniter2019}, $\Hcal(\xbm)$ corresponds to the gradient of the following regualrizer
\begin{equation}
h(\xbm) = \frac{\tau}{2}\xbm^\Tsf(\xbm-\Dcal_\sigma(\xbm)).
\end{equation}
RED has recently been theoretically analyzed for general denoisers that may not be associated with any explicit regularizer~\cite{Cohen.etal2020, Sun.etal2019c, Wu.etal2020, Sun.etal2021}.

\subsection{Other model-based deep learning}

\emph{Deep unfolding} is another widely used model-based DL methodology, originally proposed in~\cite{Gregor.LeCun2010} for sparse coding. The central idea of deep unfolding is that one can unfold an iterative algorithm and train it end-to-end as a deep neural netowork~\cite{zhang2018ista, Yang.etal2016, Hauptmann.etal2018, Adler.etal2018}.  This enables integration of the physical information into the architecture in the form of data-consistency blocks that are combined with trainable CNN regularizers~\cite{Aggarwal.etal2019, Hosseini.etal2019, Chun.etal2020, Yaman.etal2020, Aggarwal.etal2020}.
By training the corresponding model-based network end-to-end, one obtains a regularizer optimized for a specific problem.
Excellent performance of deep unfolding has been reported in a number of imaging applications~\cite{Biswas.etal2019, Hosseini.etal2019}, and recent work has addressed the computational and memory complexity of training such networks~\cite{Kellman.etal2020, Liu.etal2021}.

Another family of DL methods has used generative adversarial networks (GAN) for regularizing inverse problems~\cite{Bora.etal2017,Shah2018,Jalal.etal2020}
\begin{equation}
\label{Eq:GAN}
\zbmhat = \argmin_{\zbm\in\R^z} \|\Abm\Gcal(\zbm)-\ybm\|_2^2\quad\text{and}\quad\xbmhat=\Gcal(\zbmhat),
\end{equation}
where $\Gcal$ is a pre-trained GAN, and $\zbm\in\R^z$ is the encoding in the latent space.
The optimization in~\eqref{Eq:GAN} implicitly imposes the regularization by restricting the solution $\xbmhat$ to the range of a GAN $\Gcal$.
By searching for the optimal encoding $\zbmhat$, one can obtain an estimate $\xbmhat=\Gcal(\zbmhat)$ in the domain defined by GAN that has the smallest distance to the true $\xbm$.
The recovery properties under GANs have been analyzed in the context of compressive sensing~\cite{Bora.etal2017, Shah2018, Jalal.etal2020}.
Interested readers can find more information in the recent review~\cite{shlezinger2020model}. 

It is worth pointing out that \proposed~is complementary to all these prior works, since it seeks to learn the measurement field given measurements of a single unknown image. As shown in Section~\ref{Sec:Experiments}, \proposed~can be naturally combined with the majority of reconstruction algorithms used in computational imaging.

%%%%%%%%%%%%%%%%%%%%%%%%%%%%%%%%%%%%%%%%%%%
\begin{figure*}[t]
\begin{center}
\includegraphics[width=0.95\linewidth]{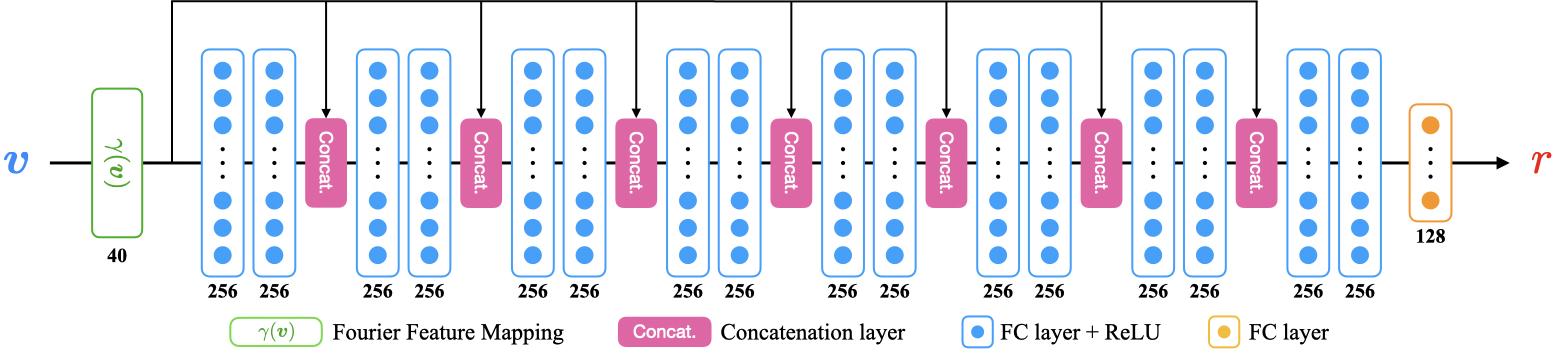}
\end{center}
\caption{Visualization of the coordinate-based MLP used in the \proposed~methodology.
The network $\Mcal_\phi = \Ncal_\phi \circ \gamma(\vbm)$ is a concatenation of a single Fourier feature mapping (FFM) layer $\gamma(\vbm)$ and a conventional MLP $\Ncal_\phi$.
As training on example pairs $\{(\vbm_i,r_i)\}_{i=1}^{N}$, $\Mcal_\phi$ is able to learn a continuous mapping from a coordinate to its response. Hence, $\Mcal_\phi$ becomes an implicit neural representation of the full-measurement field.}
\label{Fig:network}
\end{figure*}
%%%%%%%%%%%%%%%%%%%%%%%%%%%%%%%%%%%%%%%%%%%

\subsection{Deep learning for measurement synthesis}

An interest in developing DL-based approaches for measurement synthesis has recently emerged. 
An end-to-end scheme similar to image super-resolution is commonly adopted to first linearly interpolate the measurements to the same scale as that of the target measurements and then use a CNN to map the intermediate output to the final refined results~\cite{Lee.etal2017, Lee.etal2019}.
GAN has been employed to synthesize missing measurements that are corrupted by the metal artifacts~\cite{Anirudh.etal2018, Ghani.etal2019, Ghani2020, Ghani2021}.
The effectiveness of these approaches has been shown in different imaging modalities~\cite{Claus.etal2017, De.etal2019, Han.etal2020}.
Nevertheless, most deep synthesis methods require a dataset of fully-sampled measurements for supervised training.
A scan-specific CNN model that avoids training on a large dataset has been recently proposed~\cite{Mehmet.etal2018}, but still requires fully-sampled measurements of the object as groundtruth. \proposed~is fundamentally different from the existing methods for measurement synthesis since it learns a representation of the full measurement field from the measurements of an unknown object without any groundtruth.

\subsection{Deep internal learning}
\emph{Deep internal learning} explores the internal information of the test signal for learning a neural network prior without using any external data. 
One successful approach is to exploit the patch-wise similarity within images, leading to significant results for spatial and temporal super-resolution~\cite{Shocher2018,Zuckerman2020}.
Another widely adopted approach is \emph{deep image prior (DIP)}, which optimizes a CNN to parameterize the reconstructed image~\cite{Ulyanov.etal2018, Liu.etal2018a, Gandelsman2019}.
\emph{Coordinate-based neural representation} is a recent alternative that encodes a spatial field into the weights of a MLP, which is trained to map coordinates (e.g., $(x,y,z)$) to the pixel values (e.g., $[0,1]$).
It has been quite successful in computer vision and graphics, but has not been widely explored in computational imaging.
%While coordinate-based neural representation has achieved significant success in computer vision and graphics, it has not been widely explored in computational imaging. 
%This form of internal representation is beneficial because it is significantly more compact compared to the traditional processing on discrete 2D/3D grids.
The coordinate-based MLPs have been used to represent images~\cite{Tancik.etal2020}, scenes~\cite{Mildenhall.etal2020, Martin.etal2020}, and three-dimensional (3D) shapes~\cite{Chen.etal2019, Park.etal2019, Mildenhall.etal2020, Martin.etal2020}.
\emph{Neural radiance field (NeRF)}~\cite{Mildenhall.etal2020} is a recent model that has significantly improved the representation power of MLPs by first expanding the input coordinates into a Fourier spectrum (see Section~\ref{Sec:MLP} for detailed discussion).
Its formulation has been adapted for improving scene resolutions~\cite{Zhang.etal2020}, dealing with multiple lightening conditions~\cite{Srinivasan.etal2020}, removing occluders~\cite{Martin.etal2020}, and handling small deformations~\cite{Park.etal2020}.
Although there have been some early attempts in representing medical images~\cite{Tancik.etal2020}, the usage of coordinate-based representation has never been explored for representing measurement fields in computational imaging.  
This work addresses this gap by proposing \proposed~as a novel image reconstruction methodology that leverages an MLP to repesent the measurement fields.

\section{Coordinate-based Internal Learning}
\label{Sec:Proposed}
In this section, we present the details of the \proposed~methodology that leverages the power of coordinate-based learning for addressing imaging inverse problems.
Figure~\ref{Fig:workflow} illustrates the general workflow of \proposed~for a given imaging system. We first explain the proposed MLP network and then discuss its integration into several common image reconstruction methods.

\subsection{Measurement-field encoding with MLP}
\label{Sec:MLP}
The coordinate-based MLP is the central component of \proposed.
The network can be expressed as
$$\Mcal_\phi:\vbm\rightarrow r \quad\text{with}\quad\vbm\in\R^v,\;r\in\R,$$
where $\vbm$ represents the coordinate in the given imaging system, and $r$ is the corresponding sensor response.
The network can be conceptually separated into two parts, where the first part is a single \emph{Fourier feature mapping (FFM)} layer $\gamma(\vbm)$ that is pre-defined before training, while the second part is a standard MLP $\Ncal_\phi:\gamma(\vbm)\rightarrow r$ whose parameters $\phi$ needs to be optimized.  A visual illustration of the complete network architecture is provided in Figure~\ref{Fig:network}. While the numerical studies presented in this paper focus on CT, \proposed~is also applicable to other imaging modalities by simply changing the coordinate-response pairs in the MLP representation. For example, one can potentially integrate \proposed~into optical diffraction tomography (ODT)~\cite{Sung.etal2009, Kamilov.etal2015, Pham.etal2018} by letting $\vbm$ denote the sensor location and the angle of the incident light and letting $r$ have two elements corresponding to the real and imaginary components of the light-field.

\subsubsection{Fourier feature mapping} Although neural networks are known to be universal function approximators~\cite{Hornik.etal1989}, it has been found that standard MLPs perform poorly in representing high-frequency variations~\cite{Rahaman.etal2019,Mildenhall.etal2020}.
In our experiments, we also experienced such issues when we directly applied $\Ncal_\phi$ to learning the mapping $\vbm\rightarrow r$ (see \emph{No FFM} in Figure~\ref{Fig:ffm}).
In order to overcome the limitations of standard MLPs, we include the FFM layer to expand the input coordinate $\vbm$ as a combination of different frequency components
\begin{align}
\label{Eq:FFM}
\gamma(\vbm) =
%(&\sin\left(k_1\pi\vbm\right),\cos\left(k_1\pi\vbm\right),\dots,\sin\left(k_L\pi\vbm\right),\cos\left(k_L\pi\vbm\right)) \nonumber
        \begin{pmatrix}
           \sin\left(k_1\pi\vbm\right),\cos\left(k_1\pi\vbm\right), \\
           \vdots \\
           \sin\left(k_L\pi\vbm\right),\cos\left(k_L\pi\vbm\right)
         \end{pmatrix},
\end{align}
where $\sin(\cdot)$ and $\cos(\cdot)$ compute element-wise sinusoidal and cosinusoidal values, respectively, and $\{k_i\}_{i=1}^L$ determine the frequencies in the mapping.
The FFM layer pre-defines the frequency components so that the network $\Ncal_\phi$ can actively select the ones that are the most useful for encoding sensor responses by learning the weights in the first layer.
By manipulating the coefficients $k_i$ and total number of components $L>0$, we can explicitly control the expanded spectrum and thus impose some implicit regularization.
The FFM layer was first introduced in NeRF as \emph{positional encoding} of spatial coordinates~\cite{Mildenhall.etal2020}, and a follow-up work~\cite{Tancik.etal2020} has further explored its functionality by using a concept known as neural target kernels~\cite{Jacot.etal2018}.
The original formulation of $\gamma(\vbm)$ in~\cite{Mildenhall.etal2020} sets $k_i$ as an exponential function $k_i = 2^{i-1}$ with $L=10$. We discovered that the presence of very high-frequency components leads to the overfitting of the MLP to the noise in the measurements.
We thus adopted a linear sampling $k_i = \pi i/2$ in the Fourier space that leads to a higher number of frequency components in the low-frequency regions.
Our empirical results show that our strategy can effectively improve $\Mcal_\phi$ in representing high-frequency variations and preventing overfitting to noise (see Figure~\ref{Fig:ffm} for examples).

\subsubsection{MLP Architecture}
The network implementing $\Ncal_\phi$ is composed of $17$ fully-connected (FC) layers.
The first $16$ layers have $256$ hidden neurons whose outputs are activated by the rectified linear unit (ReLU), while the last layer has $128$ hidden neurons without any activation.
We implement $7$ skip connections after every even-numbered (less than $16$) FC layer to concatenate the original input of $\Ncal_\phi$ with the intermediate outputs.
The use of skip connections in MLP has been shown to be beneficial for fast training~\cite{Chen.etal2019} and better accuracy~\cite{Park.etal2019}.
Note that although $\Mcal_\phi$ is a fully connected network, its input corresponds to a single coordinate, which enables element-wise processing of all the measurements.

\subsubsection{Additional implementation details}
\proposed~trains a separate MLP to represent the full measurement field for each test objects.
This means that the training pairs $\{(\vbm_i,r_i)\}_{i=1}^{N}$ are obtained by  extracting the measurements of the test object only, without any training dataset.
The network $\Mcal_\phi$ is trained by using Adam~\cite{Kingma.Ba2015} to minimize the standard $\ell_2$-norm loss
\begin{equation}
\label{Eq:CNNLoss}
\ell(\psi)=\frac{1}{N}\sum_{i=1}^{N} \| \Mcal_{\phi}(\vbm_{i})-r_{i} \|_2^2.
\end{equation}
We implement a decreasing learning rate, which decays exponentially as the training epoch increases, to smooth our optimization.
Although $\Mcal_\phi$ is a MLP, the network has a significantly smaller size ($\approx13$ MB on disk) compared to the standard UNet architecture ($\approx108$ MB on disk) used in many DL-based models.

\subsection{Image reconstruction in \proposed}
\label{Sec:Adaption}

After training, one can generate an arbitrary number of measurements by querying $\Mcal_\phi$ using the relevant coordinates.
We will refer to the corresponding measurement field as \emph{\proposed~field}. The \proposed~field can be readily integrated into the majority of image reconstruction methods.  Here, we discuss the integration of \proposed~into four widely-used methods.

\subsubsection{Linear reconstruction}
Filtered backprojection (FBP) is a classic method for bringing the measurements into the image domain~\cite{Kak.Slaney1988}.
Since the \proposed~field is essentially a set of measurements, we can directly feed the field as input to FBP for image reconstruction.  A slightly different way to apply FBP is to form a combined input that includes both the original measurements and those generated by \proposed. The key benefit of the latter approach is that it directly uses the real data while also complementing it with \proposed~measurements.

\subsubsection{Model-based optimization}
\label{Sec:nermModelBased}
Model-based methods reconstruct images by solving optimization problems of form~\eqref{Eq:Optimization}.
The \proposed~field can be incorporated into the formulation by including an additional ``data-fidelity" term $\tilde{g}$ in the objective
\begin{equation}
\label{Eq:nermObjective}
f(\xbm) = \underbrace{(1-\alpha)g(\xbm) + \alpha\tilde{g}(\xbm)}_{\text{New data-fidelity}} + h(\xbm).
\end{equation}
Here, the parameter $0\leq\alpha\leq1$ controls the tradeoff between the real data and the generated field.
In practice, we can fine-tune the value of $\alpha$ to obtain a good balance between two terms.
For example, consider the least-squares function
\begin{equation}
\label{Eq:nermLeastSquare}
\tilde{g}(\xbm) = \frac{1}{2}\|\tilde{\Abm}\xbm-\Mcal_\phi(\tilde{\vbm})\|_2^2,
\end{equation}
where $\tilde{\Abm}\in\R^{m\times n}$ corresponds to the sampling geometry of the \proposed~field, $\tilde{\vbm}$ represents all the query coordinates for the trained MLP $\Mcal_\phi(\tilde{\vbm})$.
Since the network is pre-trained, one can directly use any existing image regularizer and solve the optimization problem with a standard iterative algorithm, such as FISTA or ADMM.

%%%%%%%%%%%%%%%%%%%%%%%%%%%%%%%%%%%%%%%%%%%
\begin{figure}[t]
\begin{center}
\includegraphics[width=0.5\linewidth]{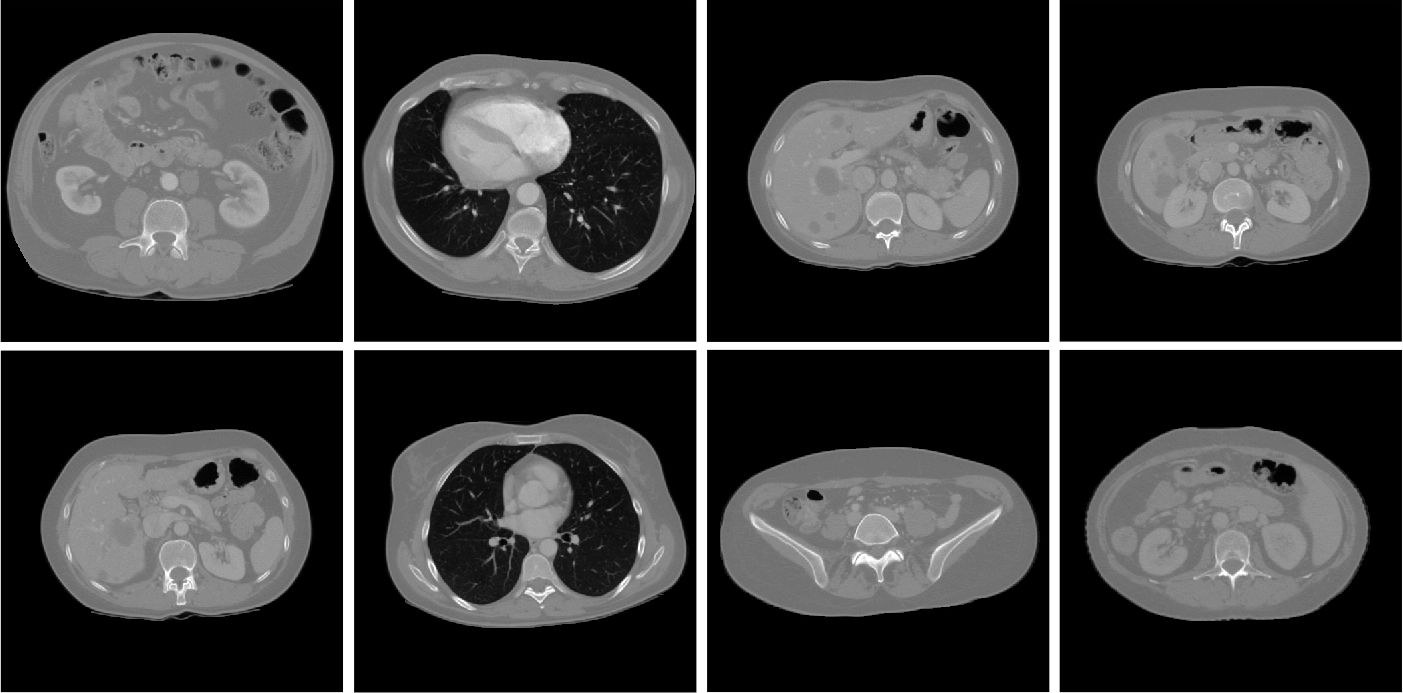}
\end{center}
\caption{Eight $512\times512$ images from the scans of two patients in the AAPM human phantom dataset~\cite{mccollough2016tu} were used for testing.}
\label{Fig:testImgs}
\end{figure}
%%%%%%%%%%%%%%%%%%%%%%%%%%%%%%%%%%%%%%%%%%%

%%%%%%%%%%%%%%%%%%%%%%%%%%%%%%%%%%%%%%%%%%%
\begin{figure}[t]
\begin{center}
\includegraphics[width=0.95\linewidth]{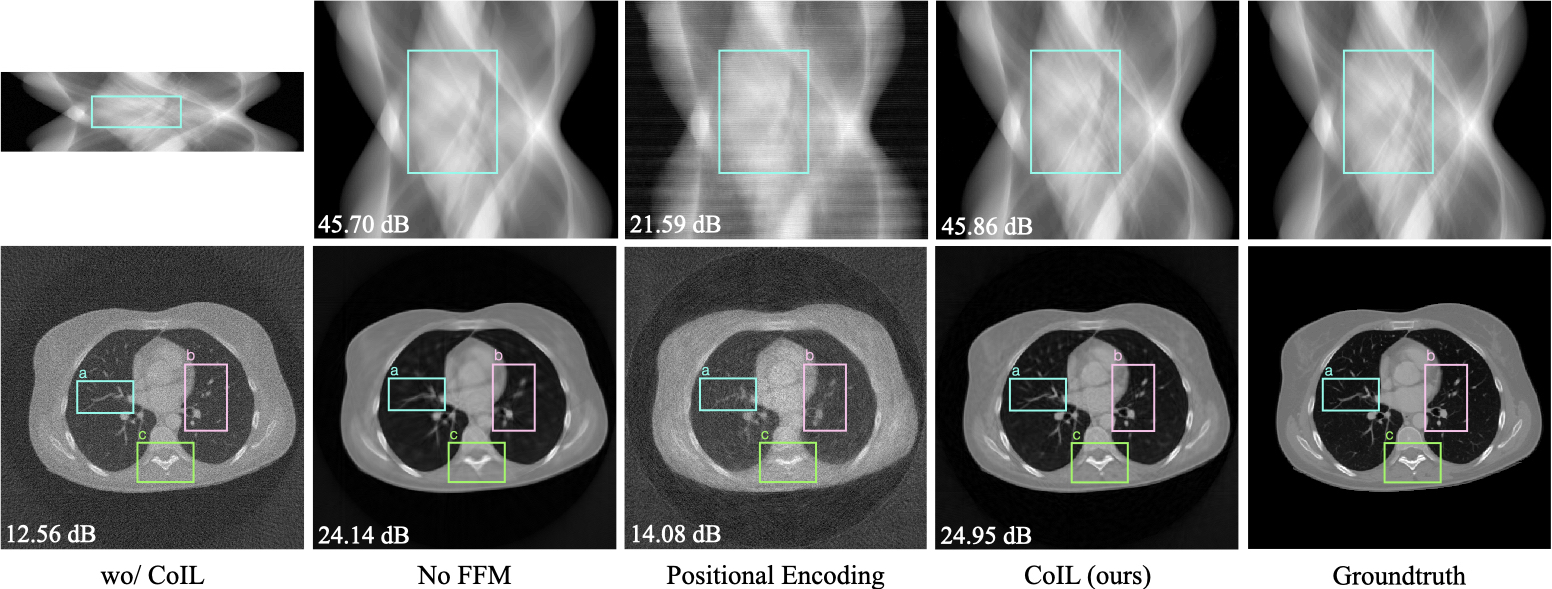}
\end{center}
\caption{Illustration of the benefit of including the Fourier feature mapping (FFM) layer into \proposed.  We plot sinograms and their FBP econstructions in the first and second row, respectively. The proposed FFM in \emph{\proposed}~is compared against \emph{No FFM} strategy, which does not have any FFM layer, and the positional encoding \emph{(Pos Enc)} adopted in~\cite{Mildenhall.etal2020}. The three MLPs are used to generate $360$ views from the $P=120$ projections with $I=40$ dB noise. Both sinograms and images are labeled with the SNR values with respect to the groundtruth shown in the right-most column. The bounding boxes highlight areas of significant visual difference. This comparison shows the benefit of using the FFM layer with linear spacing in the Fourier space.}
\label{Fig:ffm}
\end{figure}
%%%%%%%%%%%%%%%%%%%%%%%%%%%%%%%%%%%%%%%%%%%

%%%%%%%%% Table %%%%%%%%%
\begin{table}[t]
        \centering
        \caption{The average SNR of the sinograms generated by No FFM, Pos Enc, and \proposed~in the scenarios corresponding to $P\times I = \{60,90,120\}\times\{30,40,50\}$}
        \label{Tab:Sinograms}
        \vspace{5pt}
        {\footnotesize
        \begin{tabular*}{240pt}{C{30pt}C{50pt}C{35pt}C{35pt}C{35pt}}
                \toprule
                \multirow{2}{*}{\textbf{\# Views}} & \multirow{2}{*}{\textbf{Noise Level}} & \multicolumn{3}{c}{\textbf{MLP Architectures}} \\
                \cmidrule(lr){3-5}
                ($P$) & ($I$) & No FFM & Pos Enc & \proposed \\ [0.7ex]
                \midrule \\ [-2.1ex]
                \multirow{3}{*}{$60$} &$30$ & $33.95$ & $15.25$ & $\mathbf{37.34}$  \\ [0.7ex]
                & $40$ & $42.62$ & $21.79$ & $\mathbf{43.68}$ \\ [0.7ex]
                & $50$ & $46.33$ & $23.92$ & $\mathbf{48.41}$ \\ [0.7ex]
                \cdashline{1-5} \\ [-1.4ex]
                \multirow{3}{*}{$90$} &$30$ & $34.93$ & $23.68$ & $\mathbf{38.56}$ \\ [0.7ex]
                & $40$ & $43.82$ & $30.56$ & $\mathbf{44.72}$ \\ [0.7ex]
                & $50$ & $48.08$ & $35.34$ & $\mathbf{50.50}$ \\ [0.7ex]
                \cdashline{1-5} \\ [-1.4ex]
                \multirow{3}{*}{$120$} &$30$ & $36.24$ & $22.91$ & $\mathbf{39.34}$ \\ [0.7ex]
                & $40$ & $44.81$ & $24.37$ & $\mathbf{45.29}$ \\ [0.7ex]
                & $50$ & $49.68$ & $26.52$ & $\mathbf{51.59}$ \\ [0.7ex]
                \bottomrule
        \end{tabular*}}
\end{table}
%%%%%%%%%%%%%%%%%%%%%

%%%%%%%%%%%%%%%%%%%%%%%%%%%%%%%%%%%%%%%%%%%
\begin{figure}[t]
\begin{center}
\includegraphics[width=0.5\linewidth]{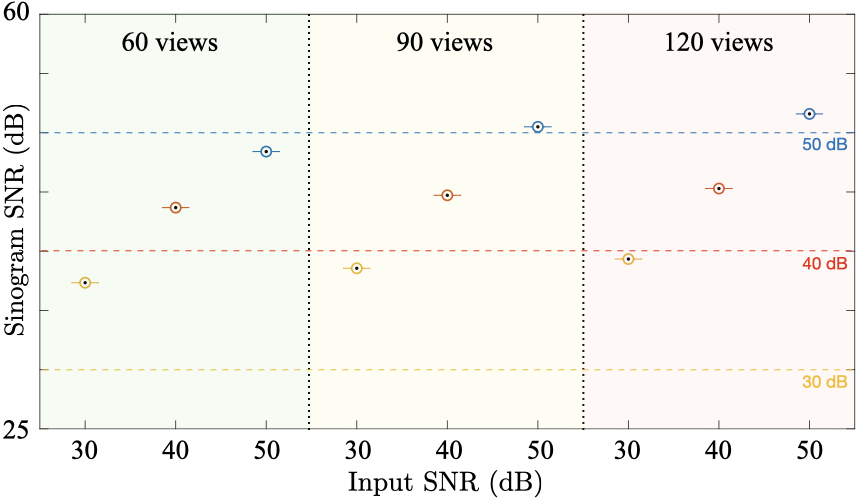}
\end{center}
\caption{Quantitative evaluation of the \proposed~field for different projection numbers ($P$) and noise levels ($I$).
The plot is divided into three regions, corresponding to $P$ equal to $60$, $90$, and $120$, respectively. Within each region, the average SNR values of the generated sinograms are plotted against different input SNR values, which are also drawn by the dotted horizontal lines for better visualization.
First, note how \proposed~generally produces measurement fields of better SNR than the noise level in the measurements. Second, the figure highlights that the quality of the generated \proposed~field improves as the number of views increases or the noise level decreases.}
\label{Fig:boxplot}
\end{figure}
%%%%%%%%%%%%%%%%%%%%%%%%%%%%%%%%%%%%%%%%%%%

\subsubsection{End-to-end DL models}
\label{Sec:nermDL}
As reviewed in Section~\ref{Sec:EndToEnd},
most end-to-end DL models are trained to directly map the low-quality images $\{\xbmtilde_i\}_{i=1}^N$ to the high-quality images $\{\xbm_i\}_{i=1}^N$, making them vulnerable to unseen outliers. For example, this adversely influences the performance of DL, when there is a mismatch between training and testing angles. \proposed~can be used to address this issue by generating the measurement field corresponding to the same subsampling rate as the measurements used for training the DL model
\begin{equation}
\label{Eq:nermDL1}
\xbmhat = \Fcal_\psi(\text{FBP}(\Mcal_\phi(\tilde{\vbm}))),
\end{equation}
where $\Fcal_\psi$ denotes the pre-trained CNN. Alternatively, one can include the original test image in the input by averaging the $\xbmtilde$ and $\text{FBP}(\Mcal_\phi(\tilde{\vbm}))$ using a weight $\alpha$
\begin{equation}
\label{Eq:nermDL2}
\xbmhat = \Fcal_\psi(\underbrace{(1-\alpha)\xbmtilde+\alpha \text{FBP}(\Mcal_\phi(\tilde{\vbm}))}_\text{Joint input}).
\end{equation}
This approach enables the usage of the learned measurements by MLP with the true measurements from the imaging system.
Our results in Section~\ref{Sec:Experiments} show that this \proposed-based strategy achieves better results than training a DL model directly on the measurements.

\subsubsection{Denoising-driven approches}

%%%%%%%%% Table %%%%%%%%%
\begin{table*}[t]
 \centering
 \caption{The average SNR values obtained with and without \proposed~by using FBP, FISTA-TV, GM-RED, and FBP-UNet in the scenarios corresponding to $P\times I = \{60,90,120\}\times\{30,40,50\}$.}
 \label{Tab:performance}
 \vspace{5pt}
{\scriptsize
 \begin{tabular*}{470pt}{C{25pt}C{50pt}C{35pt}C{35pt}C{35pt}C{35pt}C{35pt}C{35pt}C{35pt}C{35pt}C{35pt}}
  \toprule
  \multirow{2}{*}{\textbf{\# Views}} & \multirow{2}{*}{\textbf{Noise Level}} & \multicolumn{4}{c}{\textbf{without \proposed}} & \multicolumn{4}{c}{\textbf{with \proposed}} & \\
  \cmidrule(lr){3-6} \cmidrule(lr){7-10}
  ($P$) & ($I$) & FBP & FISTA-TV & GM-RED & FBP-UNet & FBP & FISTA-TV & GM-RED & FBP-UNet \\ [0.7ex]
  \midrule \\ [-2.1ex]
  \multirow{3}{*}{$60$} &$30$ & $0.15$ & $22.66$ & 22.77 & $23.44$ & $\mathbf{19.45}$ & $\mathbf{22.81}$ & $\mathbf{23.01}$ & $\mathbf{24.17}$ \\ [0.7ex]
  & $40$ & $9.09$ & $26.08$ & $27.12$  & $27.08$ & $\mathbf{23.48}$ & $\mathbf{26.95}$ & $\mathbf{27.42}$ & $\mathbf{27.93}$ \\ [0.7ex]
  & $50$ & $14.25$ & $29.37$ & $30.75$ & $29.52$ & $\mathbf{24.99}$ & $\mathbf{29.78}$ & $\mathbf{30.88}$ & $\mathbf{30.54}$ \\ [0.7ex]
  \cdashline{1-10} \\ [-1.4ex]
  \multirow{3}{*}{$90$} &$30$ & $1.95$ & $23.32$ & $23.37$ & $24.43$ & $\mathbf{20.15}$ & $\mathbf{23.58}$ & $\mathbf{23.64}$ & $\mathbf{25.20}$ \\ [0.7ex]
  & $40$ & $10.92$ & $26.98$ & $28.86$ & $28.57$ & $\mathbf{24.14}$ & $\mathbf{28.28}$ & $\mathbf{29.31}$ & $\mathbf{29.28}$ \\ [0.7ex]
  & $50$ & $16.07$ & $30.76$ & $31.71$ & $31.84$ & $\mathbf{25.69}$ & $\mathbf{31.78}$ & $\mathbf{32.19}$ & $\mathbf{32.23}$ \\ [0.7ex]
  \cdashline{1-10} \\ [-1.4ex]
  \multirow{3}{*}{$120$} &$30$ & $3.21$ & $23.79$ & $24.00$ & $24.68$ & $\mathbf{20.63}$ & $\mathbf{24.08}$ & $\mathbf{24.39}$ & $\mathbf{25.62}$ \\ [0.7ex]
  & $40$ & $12.10$ & $27.59$ & $29.30$  & $29.18$ & $\mathbf{24.52}$ & $\mathbf{28.95}$ & $\mathbf{29.79}$ & $\mathbf{29.71}$ \\ [0.7ex]
  & $50$ & $17.07$ & $31.53$ & $32.31$ & $32.80$ & $\mathbf{26.02}$ & $\mathbf{32.89}$ & $\mathbf{33.02}$ & $\mathbf{33.31}$ \\ [0.7ex]
  \bottomrule
 \end{tabular*}}
\end{table*}
%%%%%%%%%%%%%%%%%%%%%

%%%%%%%%%%%%%%%%%%%%%%%%%%%%%%%%%%%%%%%%%%%
\begin{figure}[t]
\begin{center}
\includegraphics[width=0.5\linewidth]{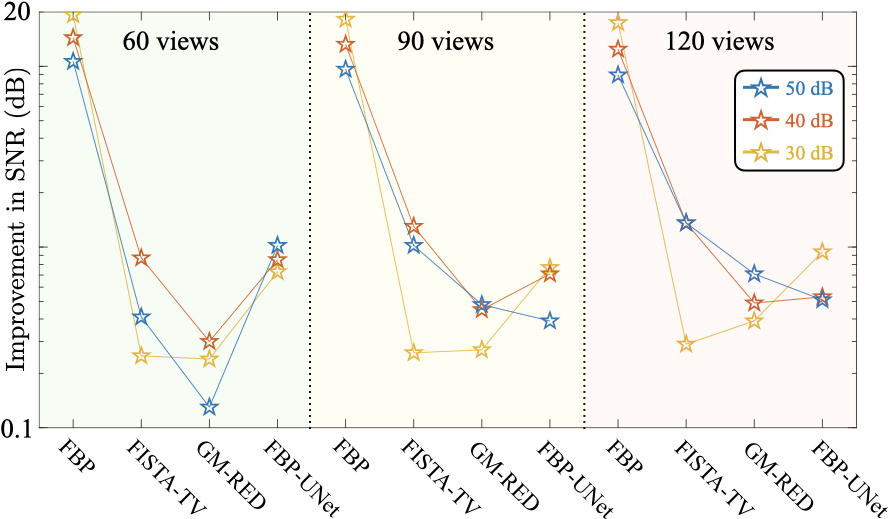}
\end{center}
\caption{SNR improvements due to \proposed~for each reconstruction algorithm. The plot is divided into three regions, corresponding to $60$, $90$, and $120$ projections, respectively. Within each region, the average SNR improvement is plotted against the reconstruction method.
The vertical axis is in log-scale for better visualization. Note that \proposed~consistently improves the average SNR values for all the considered algorithms in every scenario.}
\label{Fig:improvement}
\end{figure}
%%%%%%%%%%%%%%%%%%%%%%%%%%%%%%%%%%%%%%%%%%%

PnP/RED algorithms can be interpreted as extensions of model-based algorithms balancing consistency with the measurements against deep denoising priors~\cite{Sun.etal2019c,Ryu.etal2019}.
Consider gradient-based RED (GM-RED)
\begin{equation}
\label{Eq:GM-RED}
\xbm^+\leftarrow\xbm-\gamma\left[\nabla g(\xbm) + \tau(\xbm-\Dcal_\sigma(\xbm))\right]
\end{equation}
where $\gamma>0$ is the stepsize, and $\nabla g$ is the gradient of the data-fidelity term. Similar to the modification of model-based optimization, one straightforward way to integrate \proposed~into GM-RED is to include the gradient of $\tilde{g}$ as an extra term
\begin{equation*}
%\label{Eq:nermGM-RED}
\xbm^+\leftarrow\xbm-\gamma[\underbrace{(1-\alpha)\nabla g(\xbm) + \alpha\nabla \tilde{g}(\xbm)}_{\text{New data enforcement}} + \tau(\xbm-\Dcal_\sigma(\xbm))],
\end{equation*}
where the new update ensures the consistency with the real measurements as well as the field generated by \proposed, with $\alpha$ controlling the relative weighting. This idea is also applicable to PnP, for example, by integrating \proposed~within PnP-FISTA
\begin{subequations}
\begin{align}
\xbm^+ &\leftarrow \Dcal_\sigma(\sbm-\gamma[(1-\alpha)\nabla g(\sbm) + \alpha\nabla \tilde{g}(\sbm)]) \\
\sbm^+ &\leftarrow \xbm^+ + ((q^+-1)/q^+)(\xbm^+-\xbm)
\end{align}
\end{subequations}
where the acceleration parameter $q>0$ is updated as
$$q^+\leftarrow\frac{1}{2}\left(1+\sqrt{1+4q^2}\right).$$
In the next section, we will provide results highlighting the performance of \proposed~in the context of all these algorithms.

\section{Numerical Validations}
\label{Sec:Experiments}

We numerically validate \proposed~in the context of sparse-view CT.
We first substantiate the effectiveness of the proposed form of FFM, and then demonstrate the benefits of using \proposed~for image reconstruction.
We consider four reconstruction methods, \emph{FBP}, \emph{FISTA-TV}, \emph{GM-RED}, and \emph{FBP-UNet}.
FBP-UNet refers to the end-to-end model proposed in~\cite{DJin.etal2017} and FISTA-TV refers to the TV regularized inversion implemented using FISTA.
We integrate \proposed~into these algorithm by including the parameter $\alpha$ as discussed in Section~\ref{Sec:Adaption}.

%%%%%%%%%%%%%%%%%%%%%%%%%%%%%%%%%%%%%%%%%%%
\begin{figure*}[t]
\begin{center}
\includegraphics[width=0.95\linewidth]{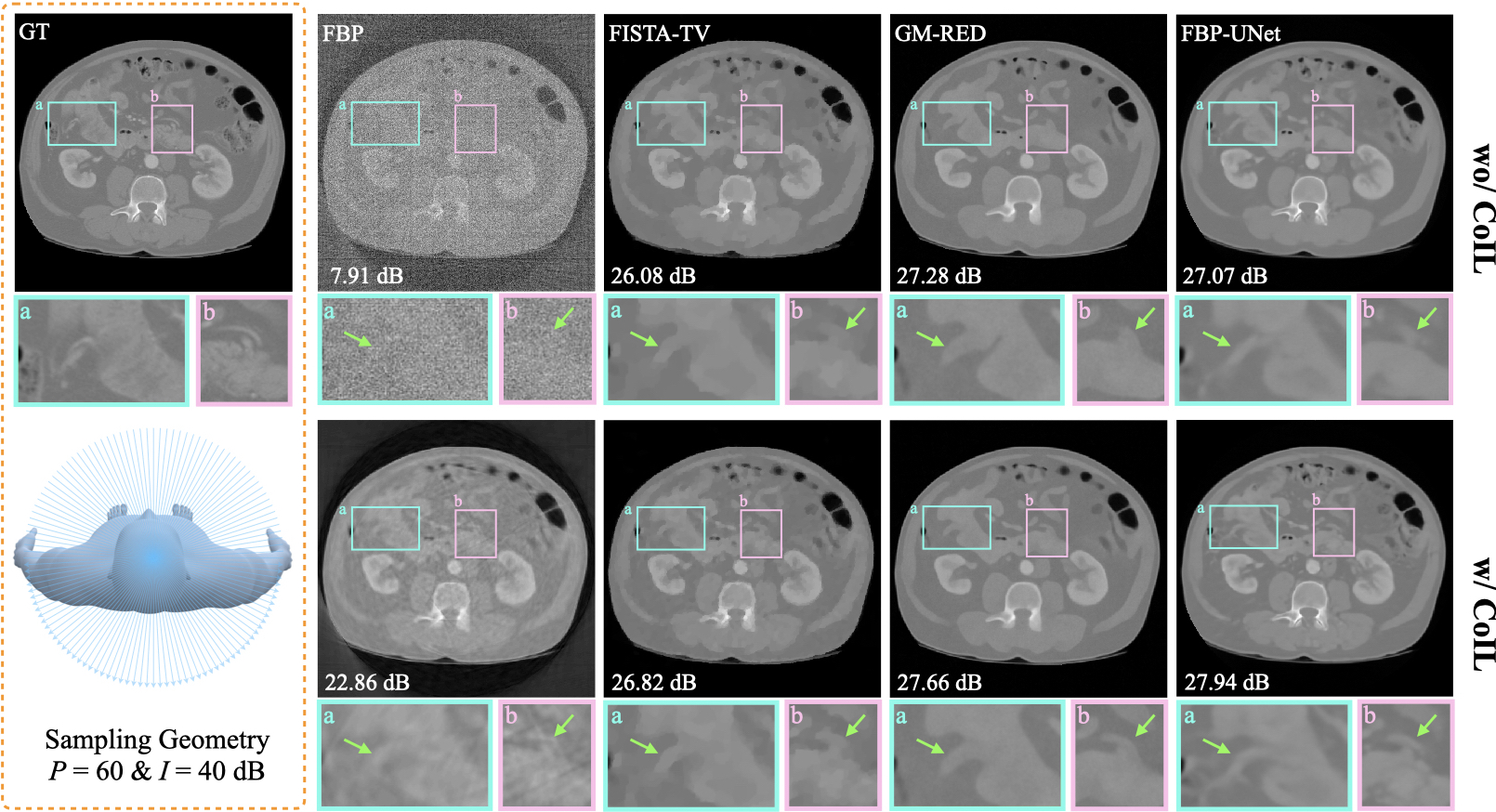}
\end{center}
\caption{Visual illustration of reconstruction with and without \proposed~using the several methods. \proposed~generates measurement fields corresponding to $360$ (for FBP, TV, and RED) and $90$ (used for FBP-UNet) views from $P=60$ measurements with $I=40$ dB noise. Each image is labeled with its SNR value with respect to the groundtruth displayed in the left-most column. The visual differences are highlighted in the bounding boxes using green arrows. Note how \proposed~enables the recovery of certain details missing in the reconstructions without it.}
\label{Fig:performance1}
\end{figure*}
%%%%%%%%%%%%%%%%%%%%%%%%%%%%%%%%%%%%%%%%%%%

%%%%%%%%%%%%%%%%%%%%%%%%%%%%%%%%%%%%%%%%%%%
\begin{figure*}[t]
\begin{center}
\includegraphics[width=0.95\linewidth]{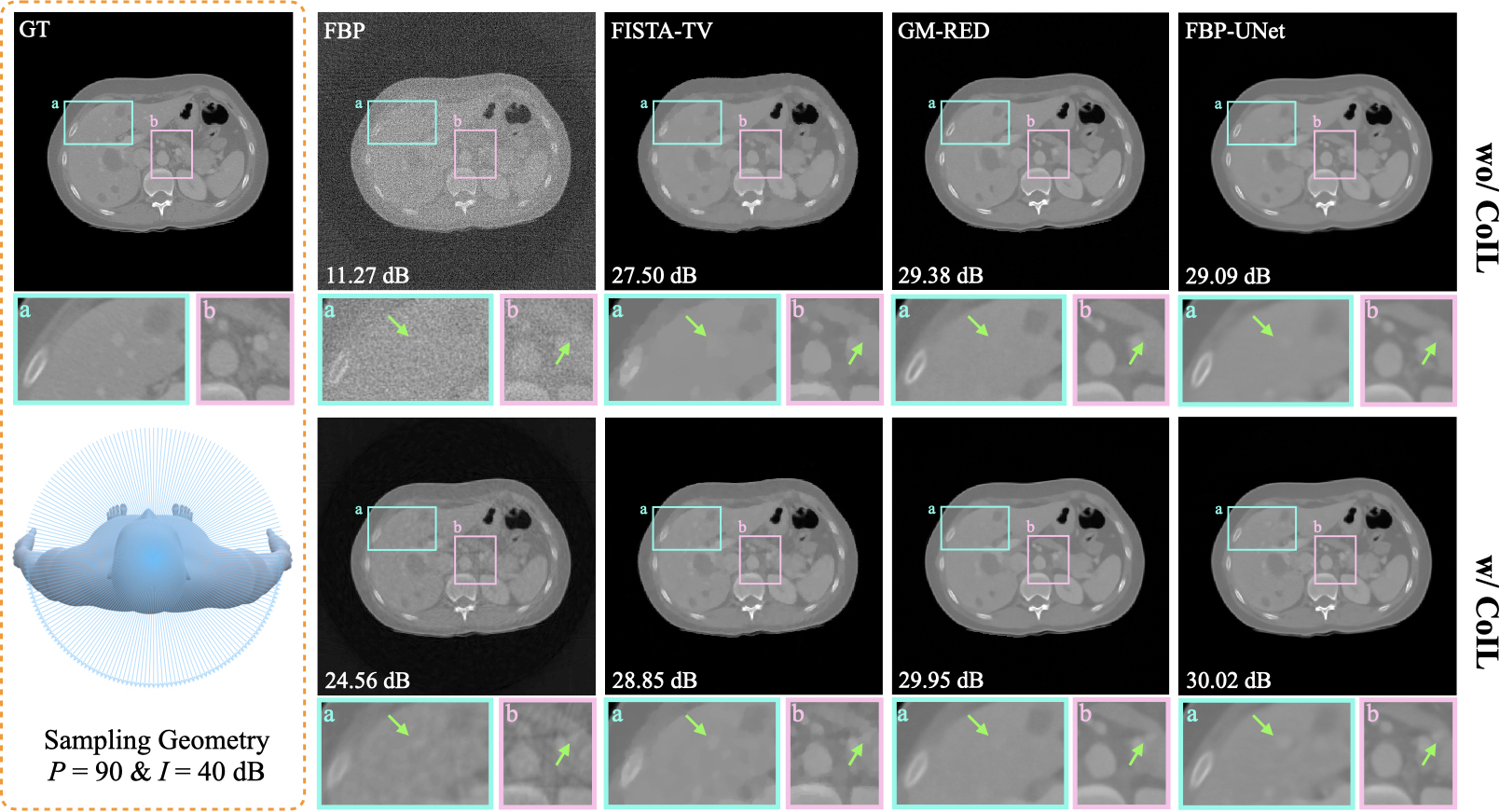}
\end{center}
\caption{Visual illustration of reconstruction with and without \proposed~using the several methods. \proposed~generates measurement fields corresponding to $360$ (for FBP, TV, and RED) and $135$ (used for FBP-UNet) views from $P=90$ measurements with $I=40$ dB noise. Each image is labeled with its SNR value with respect to the groundtruth displayed in the left-most column. The visual differences are highlighted in the bounding boxes using green arrows.}
\label{Fig:performance2}
\end{figure*}
%%%%%%%%%%%%%%%%%%%%%%%%%%%%%%%%%%%%%%%%%%%

%%%%%%%%%%%%%%%%%%%%%%%%%%%%%%%%%%%%%%%%%%%
\begin{figure*}[t]
\begin{center}
\includegraphics[width=0.95\linewidth]{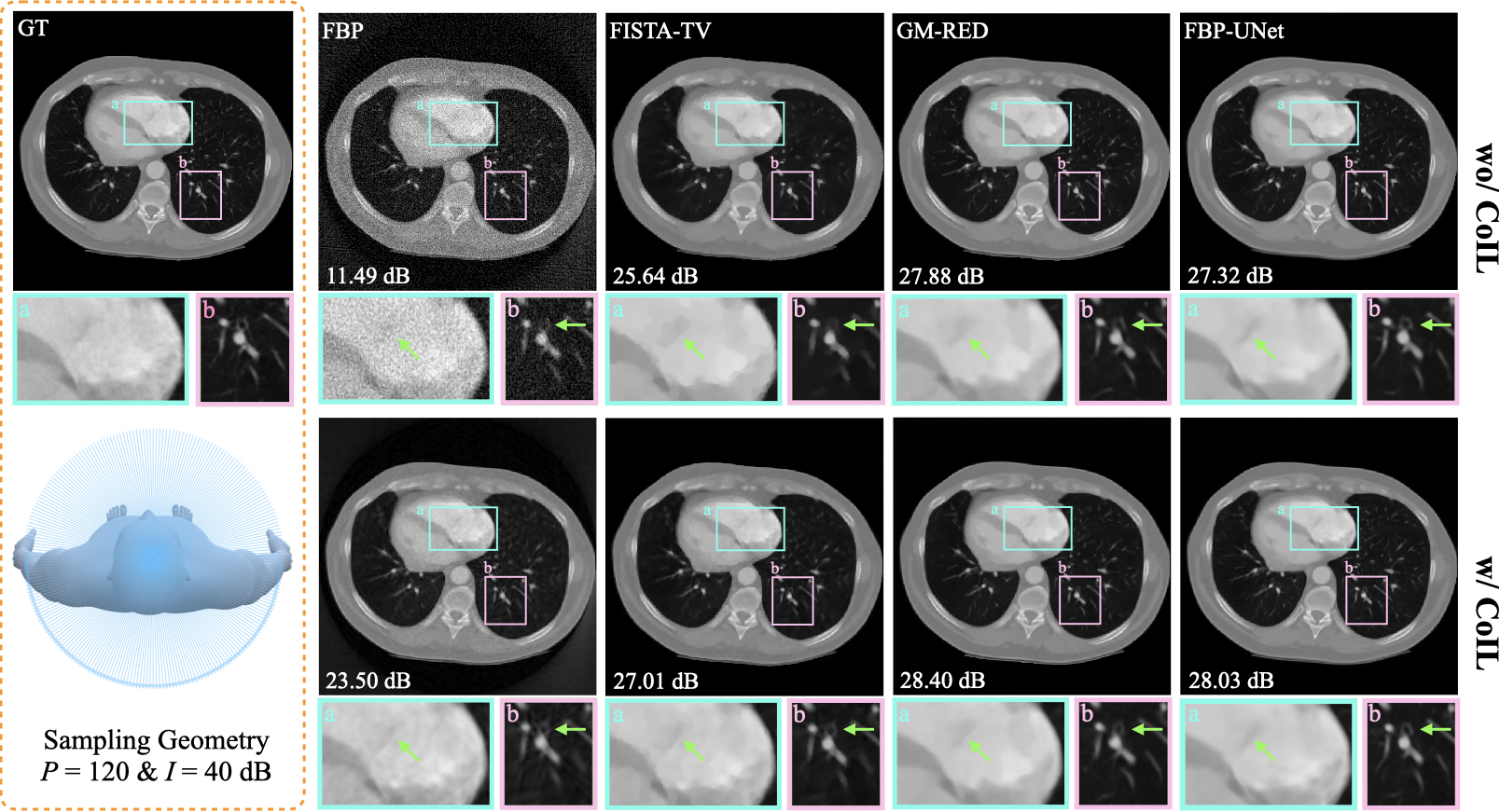}
\end{center}
\caption{Visual illustration of reconstruction with and without \proposed~using the several methods. \proposed~generates measurement fields corresponding to $360$ (for FBP, TV, and RED) and $180$ (used for FBP-UNet) views from $P=120$ measurements with $I=40$ dB noise. Each image is labeled with its SNR value with respect to the groundtruth displayed in the left-most column. The visual differences are highlighted in the bounding boxes using green arrows.
Visual examples reconstructed with and without \proposed~using the considered methods.}
\label{Fig:performance3}
\end{figure*}
%%%%%%%%%%%%%%%%%%%%%%%%%%%%%%%%%%%%%%%%%%%

\subsection{Sparse view CT and experimental setup}
Sparse view X-ray CT is an imaging modality that aims to reconstruct a tomographic image from few projections.
In medical applications, it can significantly reduce the radiation dose and hence reduce the risk of radiation exposure.
The reconstruction task in CT can be formulated as the linear inverse problem of form~\eqref{Eq:inverseProblem}.
In our simulations, we adopt the parallel beam geometry with a measurement operator $\Abm$ corresponding to the Radon transform.

We consider the experimental setting where the X-ray beam is emitted from the view angle $\theta\in[0,\pi]$ and its radiation attenuation is recorded by the detectors at different (normalized) sensor-plane locations $l\in[0,1]$.
Thus, the MLP is trained to map the location and angle $(l,\theta)$ to the corresponding response $r$. Figure~\ref{Fig:testImgs} visualizes eight $512\times512$ test images used in all experiments.
These images are selected from the scans of two patients in the APPM human phantom dataset\footnote{The 2016
NIH-AAPM-Mayo Clinic Low Dose CT Grand Challenge}~\cite{mccollough2016tu}, while the scans of other patients are used for training the FBP-UNet and the deep denoiser in GM-RED.
We implemented $\Abm$ and its adjoint $\Abm^{\Tsf}$ by using \texttt{RayTransform} from the Operator Discretization Library (ODL)~\cite{adler.etal2017}, which allows fast computation using a GPU backend.
We synthesized the test sinograms corresponding to $P\in\{60,90,120\}$ projection views, each further contaminated by three noise levels equivalent to the input signal-to-noise ratio (SNR) of $I\in\{30,40, 50\}$ dB.
SNR is also used as a metric to quantify the reconstruction quality
\begin{align}
\operatorname{SNR}(\hat{\xbm}, \xbm) \triangleq20 \operatorname{log} _{10}\left(\frac{\|\xbm\|_2}{\|\xbm-\hat{\xbm}\|_2}\right).
\end{align}
We denote the SNR values averaged over all test images as \emph{average SNR}.

For each test image, \proposed~trains separate MLPs to represent its full measurement field in different scenarios $P\times I = \{60,90,120\}\times\{30,40,50\}$. We conducted all the experiments, as well as the training of all neural networks, on a machine equipped with an Intel Xeon Gold 6130 Processor and four Nvidia GeForce GTX 1080 Ti GPUs. The training time of a single MLP on our machine takes about $30$ minutes.

\subsection{Effectiveness of the FFM layer}
We first evaluate the effectiveness of the proposed FFM layer used in the coordinate-based MLP.
We trained and compared three networks where: (a) the FFM layer is not implemented (\emph{No FFM}); (b) the FFM layer implements the positional encoding where $k_i=2^{k-i}$ (\emph{Pos Enc}); and (c) the FFM layer implements the proposed linear expansion $k_i=(\pi i)/2$ (\emph{\proposed}).
In the simulations, we use these networks to generate the sinograms corresponding to $360$ views, and the total number of frequency components is set to $L=10$ for both Pos Enc and \proposed.

Table~\ref{Tab:Sinograms} summarizes the average SNR values of the sinograms generated by the three networks in all scenarios.
Here, we use SNR as the quality metric in the sinogram space, because it enables straightforward comparison with the original measurements whose noise level is characterized by input SNR.
As shown in the table, \proposed~consistently achieves significantly higher SNR values than both \emph{No FFM} and \emph{Pos Enc}.
Our interpretation is that \emph{No FFM} is unable to represent the high-frequency variations in the measurement field, while \emph{Pos Enc} overfits to noise by containing too many high-frequency components.
We observe that the sampling pattern in \emph{\proposed}~better captures the nature of the measurements without overfitting to the noise. 
This is further illustrated in visual examples in Figure~\ref{Fig:ffm}, which plots the sinograms and their FBP reconstructions obtained by each network for $P = 120$ and $I=40$ dB.
Specifically, \emph{No FFM} is able to represent the general structure of the sinogram but fails in generating the details; \emph{Pos Enc} produces strong artifacts in its sinogram due to its FFM layer. 
\emph{\proposed}~succeeds in both representing the high-frequency details and avoiding strong artifacts in the generated measurements.
The improvement in the sinogram quality is also reflected in the SNR values obtained after FBP reconstruction. 
Note how \emph{\proposed}~significantly differs from other approaches in the regions highlighted by the bounding boxes.

We have also investigated the evolution of the sinogram quality for different number of views and noise levels. Figure~\ref{Fig:ffm} plots the SNR of the sinograms obtained by \proposed~against the input SNR ($I\in\{30,40,50\}$) for different number of views ($P\in\{60,90,120\}$).
The three dotted horizontal lines in the figure highlight each $I$ value.
We first note that \proposed~generates sinograms that generally have higher SNR than the noise level in the measurements.
In particular, when $I=30$ dB, the average SNR values are more than $7$ dB higher for every $P$. This highlights the ability of \proposed~to generate high-quality sinograms. The figure also demonstrates that the SNR values improve as the number of views increases or noise level decreases. This highlights that the quality of the \proposed~fields can be improved by having more measurements or acquiring those that are less noisy.

\subsection{Performance evaluation of \proposed}
We next highlight the benefit of \proposed~for image reconstruction.
We trained all our MLPs by using the FFM layer based on our linear expansion.
We implemented FBP by using \texttt{fbp-op} from the ODL package under the default parameter setting.
We used DnCNN~\cite{Zhang.etal2017} to build the deep denoiser within GM-RED. 
In every experiment, we selected the network achieving the highest SNR value from the ones corresponding to five noise levels $\sigma\in\{5, 10, 15, 20, 25\}$.
For FBP, FISTA-TV, and GM-RED, \proposed~generates the measurement field with $360$ projection views from the test measurements. For FBP-UNet using \proposed, we trained the CNN on the dataset consisting of the measurements having $1.5\times P=\{90,135,180\}$ projection views and used \proposed~to generate additional measurements to achieve that number. As a baseline, we trained a separate FBP-UNet  that directly predicts the groundtruth form the test measurements.  Note that the baseline networks correspond to the optimal performance that FBP-UNet can achieve for the test measurements without integrating \proposed.
In order to stabilize FBP-UNet, we trained these networks using the data with random fluctuations in both projection views ($\pm15$) and noise amount ($\pm 5$ dB).

Figure~\ref{Fig:improvement} quantitatively evaluates the improvements in imaging quality due to \proposed~for all the considered reconstruction algorithms. For each algorithm, we plot the difference between the SNR obtained with and without \proposed. We can clearly observe that \proposed~leads to significant quality improvements for all the algorithms. Remarkably, for the higher amount of noise ($I=30$ dB), the average improvement by \proposed~can sometimes be as high as $20$ dB for FBP. On the other hand, when the noise is relatively mild, \proposed~still leads to significant SNR improvements for all algorithms including model-based and DL-based methods. In particular, for $P=60$ and $I=50$ dB, FBP-UNet without \proposed~achieves $29.52$ dB, while FBP-UNet with \proposed~achieves $30.54$ dB, which is nearly $1$ dB improvement.
The exact numbers obtained by each algorithm are also summarized in Table~\ref{Tab:performance}. These results highlight that \proposed~is able to accurately represent the measurement field and generate high-fidelity measurements that can be used to improve image reconstruction.

Figure~\ref{Fig:performance1} presents visual comparisons of images reconstructed with and without \proposed~for $P=60$ and $I=40$.
Each image is labeled with its SNR with respect to the groundtruth and the visual differences are highlighted by arrows in the bounding boxes. This comparison highlights visual improvements due to \proposed. For example, consider the visual differences for FBP-Unet, where one can clearly see additional visual details after integration of the \proposed~field. The better reconstruction quality  with \proposed~is also reflected in the higher SNR values. Additional visual comparisons in Figure~\ref{Fig:performance2} and Figure~\ref{Fig:performance3} also highlight the benefit of image reconstruction with \proposed.

\section{Conclusion}
\label{Sec:Conclusion}
The \proposed~methodology developed in this paper is a new approach for computational imaging using coordinate-based neural representations.  \proposed~can represent the full measurement field as a single MLP network by training it to map the measurement coordinates to their sensor responses.  This makes \proposed~a self-supervised model that can be trained without any external dataset.  We provided extensive empirical results demonstrating the improvements due to \proposed~in the context of sparse-view CT, highlighting its great potential to work synergistically with existing image reconstruction methods.  Future work will explore new applications of \proposed~to other imaging modalities, such as optical diffraction tomography and intensity diffraction tomography.

\section*{Acknowledgement}
The research in this work was supported by NSF award CCF-1813910 and by the Laboratory Directed Research and Development program of Los Alamos National Laboratory under project number 20200061DR.

\end{document}